# Implications of magnetic and magnetodielectric behavior of GdCrTiO$_5$

Tathamay Basu, Kiran Singh,[@] Smita Gohil, Shankar Ghosh, and E.V. Sampathkumaran

*Tata Institute of Fundamental Research, Homi Bhabha Road, Colaba, Mumbai-400005, India*

[@]Present address: UGC-DAE Consortium for Scientific Research, University Campus, Khandwa Road, Indore-452001, India

**Abstract**

We have carried out dc magnetization (*M*), heat-capacity (*C*) and dielectric studies down to 2K for the compound GdCrTiO$_5$, crystallizing in orthorhombic *Pbam* structure, in which well-known multiferroics RMn$_2$O$_5$ (R= Rare-earths) form. The points of emphasis are: (i) The magnetic ordering temperature of Cr appears to be suppressed compared to that in isostructural Nd counterpart, NdCrTiO$_5$. This finding on the Gd compound suggests that Nd 4f hybridization plays an uncommon role on the magnetism of Cr in contrast to a proposal long ago. (ii) Dielectric constant does not exhibit any notable feature below about 30 K in the absence of external magnetic field, but a peak appears and gets stronger with the application of external magnetic fields, supporting the existence of magnetodielectric coupling. (iii) The dielectric anomalies appear even near 100 K, which can be attributed to short-range magnetic-order. We also observe a gain in spectral weight below about 150 K in Raman spectra in the frequency range 150 to 400 cm$^{-1}$, which could be magnetic in origin supporting short-range magnetic order. It is of interest to explore whether geometrically frustration plays any role on the dielectric properties of this family, as in the case of RMn$_2$O$_5$.

The rare-earth ($R$) compounds of the type $R$Mn$_2$O$_5$, crystallizing in an orthorhombic structure (space group *Pbam*), have attracted a lot of attention with respect to magnetoelectric (ME) coupling and magnetic-field induced ferroelectric behaviour [see references 1-4 and articles cited therein]. Very recently, possibility of the ferroelectric phase of this compound even at room temperature has been demonstrated[5]. However, the members of the family $R$CrTiO$_5$ belonging to the same structure were paid very little attention in this field, except some work on $R$= Nd. Interestingly enough, this Nd compound was considered to be one of the first few ME materials with two distinct magnetic sublattices at that time of investigation, nearly three decades[6,7]. It was argued that Cr ions order cooperatively antiferromagnetically at ($T_N=$) 21 K, which in turn triggers magnetic ordering of Nd sublattice at a lower temperature ($T \leq 13K$)[7]. There was no further work in the literature to address this issue. This system was subjected to further extensive ME investigations recently by Hwang et al[8] and there was some ambiguity whether this compound can be classified as 'multiferroic' in view of nonobservation of electric-polarization in the absence of an externally applied magnetic-field ($H$), though ME coupling sets in at $T_N$. In fact, such a class of materials exhibiting magnetic-field-induced effects, e.g., MnTiO$_3$, generated enough interest in this field [see, for example, Refs. 9, 10] and therefore the identification of such materials is of great interest. Subsequent work by Saha et al[11] addressed this issue further and concluded that this compound is a genuine multiferroic material, possibly driven by collinear magnetostriction. Vopson et al[12] proposed that this compound is an example for their theory of 'multicaloric effect'. Thus, the efforts to understand the properties of this compound and this family is scarce. We think that it is important to investigate other members of this rare-earth series for magnetic and dielectric properties, primarily to bring to the notice of the scientific community the existence of another family of magnetodilectric compounds in the same structure as RMn$_2$O$_5$. With this in mind, we have investigated the isostructural compound GdCrTiO$_5$ by *dc* magnetization ($M$), heat capacity ($C$), complex dielectric measurements and Raman spectroscopy. The results point towards a possible role of (i) Nd 4f hybridization due to (well-known) large 4f radial extension on the magnetic and ME behavior of the Nd analogue and (ii) short-range electric and magnetic correlations persisting over a wide temperature range, possibly resulting in nanopolar region.

Polycrystalline specimen of $GdCrTiO_5$ was prepared by solid state reaction method as described by Hwang et al[8] starting with stoichiometric amounts of high-purity (>99.95%) oxides, $Gd_2O_3$, $Cr_2O_3$ and $TiO_2$. X-ray diffraction (XRD) study (including Rietveld refinement, see figure 1) established that the desired compound is formed in single phase with proper structure (in *Pbam* space group) within the resolution (<2%) of the diffractometer. The lattice parameters are found to be *a*= 7.4172 (8) Å, *b*= 5.6664 (4) Å, and *c*= 5.7688 (4) Å. The *dc M* was measured as a function of *T* in several magnetic fields and also as a function of *H* at selected temperatures with the help of a commercial superconducting quantum interference device (Quantum Design, QD) magnetometer. Heat-capacity behaviour was tracked in the presence of external magnetic fields with a commercial Physical Properties Measurement System (PPMS, QD). The complex dielectric measurements were carried out using a LCR meter (Agilent E4980A) with a homemade sample holder integrated with the PPMS, after cooling the sample in the absence of an external magnetic-field; such measurements were carried out not only in the absence of *H*, but also in the presence of various *dc* magnetic fields. The Raman spectra were collected using a triple grating Raman spectrometer (T64000: Jobin Yvon) equipped with a liquid nitrogen cooled charge-coupled device in the back scattering geometry. The excitation source was the 514.5 nm line of a mixed gas laser (Stabilite 2018: Spectra Physics). A continuous flow helium cryostat (Microstat He: Oxford Instruments) was used to carry out the low temperature Raman measurements.

Temperature dependence of *dc* susceptibility ($\chi$) measured in various fields is shown in figure 2a. It is obvious that there is a monotonous increase of $\chi$ with decreasing *T*. Inverse $\chi$, shown for *H*= 5 kOe in figure 2b, is found to be linear above 150 K. From the Curie-Weiss fitting of the data from the linear region (say, above 200 K), the values of the paramagnetic Curie temperature ($\theta_P$) and effective moment ($\mu_{eff}$) were obtained, which are found to be ~ -25K and 8.54 $\mu_B$ per formula unit, respectively. The negative sign of $\theta_P$ suggests dominant antiferromagnetic interactions in this compound. Assuming theoretical paramagnetic moment of Gd moment (7.94 $\mu_B$), the deduced value of $\mu_{eff}$ on Cr is found to be 3.15 $\mu_B$ which is less than the spin only (S=3/2) moment value of $Cr^{+3}$ (3.87 $\mu_B$). Possibly there is a disorder in the lattice[7,8] as a consequence of which, for some Cr ions, the magnetic moment is lower. No well-defined anomaly that is attributable to magnetic ordering is transparent in the plot of $\chi$ versus *T*. However, a careful look at the derivate plot of $\chi$ under 100 Oe (see right inset of figure 2a) reveals a sudden change of slope around 10 K, which could be due to magnetic ordering from Cr. The derivate plot of $\chi$ measured in 70 kOe also exhibits anomalies around 7-10 K, as shown in the left inset of figure 2a. It appears that this magnetic ordering does not result from Gd. We did not find any difference in the curves obtained for zero-field-cooling (ZFC) and field-cooling (FC) conditions of the specimen and therefore spin-glass freezing can be ruled out. The fact that Gd remains paramagnetic down to (at least) 5 K can be inferred from the plot of isothermal $M$ at 5 K, the shape of which follows closely the Brillouin function (see figure 2c). The value of saturation magnetization at the highest field employed is only marginally higher compared to the free-ion value for trivalent Gd. It therefore appears that the magnetic-moment contribution from Cr in the magnetically ordered state is very small (compared to the value expected for $S=3/2$), thereby implying that Cr ions, if ordered magnetically, should be antiferromagnetic.

We would like to emphasize that the deviation from Curie-Weiss behavior occurs at high temperatures (below 200 K), whereas any long-range magnetic ordering from Gd or Cr, if present, does not seem to occur above 15 K (see also below). Distinct $H$-dependence of $\chi(T)$ even around 120 K is also seen, as shown in the inset of figure 2b by comparison of the curves for $H=$ 100 Oe and 5 kOe. Crystal-fields effects can not be the origin of such effects, as Gd is a S-state ion. Therefore, short-range magnetic correlations persisting over a wide $T$-range should be responsible for this magnetic behavior.

In order to get further insight about magnetic ordering, we show the plot of $C$ versus $T$ in figure 3a and a broad (but feeble) peak appears around 10 K, supporting the existence of an anomaly. Since the anomaly is weak, clearly, the entropy associated with possible magnetic ordering must be very small. There is an upturn below about 8 K in $C$ versus $T$. This signals that there is another magnetic peak below 2 K, possibly arising from Gd. In order to gain further knowledge about the nature of magnetism of Gd, we have obtained $C(T)$ plots with several fields. As shown in figure 3b, an application of magnetic field suppresses this upturn and a drop (superimposed over lattice contribution) appears with lowering temperature. The temperature at which this drop occurs increases with $H$, appearing near 8 K for 80 kOe. This upward shift of this temperature with $H$ suggests that the ordering is of a ferromagnetic-type below 2 K in zero field. In order to gain information about short-range magnetic correlations, we show the data obtained in several fields, over a wider $T$-range in the form of $C/T$ versus $T$, in figure 3c. It is obvious from this figure that the curves tend to deviate from each other below about 100 K. This appears to support our conclusion above on the existence of short-range magnetic order. It is surprising to note that the absolute values of $C/T$ at a given temperature in the region of proposed short-range order (10 to 100 K) increases with $H$, rather than suppressing the values, and the exact origin of this is not clear to us at this moment. We speculate that this is due to an interplay with an additional phenomenon, say, formation of nanopolar regions proposed below.

$T$-dependencies of dielectric constant ($\varepsilon'$) and loss factor (tan$\delta$) at low temperatures (<40 K) are shown in figure 4a and 4b for a fixed frequency ($\nu$) of 50 kHz in different magnetic fields. No clear feature is observed below 40 K in $\varepsilon'(T)$ in the absence of a magnetic field. However, a well-defined peak gradually builds up around 10 K with the application of $H$. Correspondingly, there is a drop in tan$\delta$, even for $H= 0$, which is also $H$-dependent. These features clearly reveal the presence of magnetodieletric (MDE) coupling in this compound. It is remarkable that absence of any prominent feature in zero field and notable MDE behavior are similar to those known for $NdCrTiO_5$[8] and $MnTiO_3$[9]. We have also taken data at several frequencies (1 - 100 kHz) and we could not resolve any $\nu$-dependence of the peak temperature and hence glassy nature of this feature is excluded. Here we want to mention that this system is highly insulating, which is apparent from the very low value (< 0.001) of the loss part and therefore the observed features are intrinsic. The resistance of the sample falls in gega-ohm range at room temperature and increases sharply with lowering temperature; the dc resistivity cannot be measured below 200 K as it attains tera-ohm range. While the weak heat-capacity peak around 10 K (mentioned earlier) could be magnetic in origin (possibly from Cr), we can not rule out that this $C(T)$ feature arises from these electric dipole anomalies. We would also like to add that we attempted to detect spontaneous electric polarization around 10 K, but the feature is found to be so weak that we are not able to attach any significance confidently at this moment.

In order to confirm the existence of MDE, we have measured dielectric constant as a function of $H$ at different temperatures (see figure 5). The change in dielectric constant with varying $H$ is maximum in the vicinity of the dielectric peak in $\varepsilon'(T)$, which reduces with increasing or decreasing temperature, however remaining finite even in the range 50 - 100 K. The MDE coupling vanishes at further higher temperatures. This result suggests the existence of MDE coupling due to short range magnetic correlations.

We have made some interesting observations in the dielectric data at higher temperatures (>>50 K) as well. The curves obtained in a wider temperature range (below 250 K) at several $\nu$ are shown in figure 6 in the absence of any external magnetic field. It is distinctly clear that the absolute values of $\varepsilon'$ (figure 6a) and tan$\delta$ (figure 6b) are strongly $\nu$-dependent, essentially showing the tendency to get suppressed with $\nu$. In this system, the change in dielectric constant with $T$ is also very small, a useful condition for capacitor applications. Additionally, there is a shoulder in ε′ and a tendency for a peak in tanδ around 100 K, for instance, for 10 kHz, and these features move dramatically to a higher-$T$ with increasing ν (for instance to about 150 K for 100 kHz). Such a strong dispersion characteristic of relaxor ferroelectrics can be attributed to polar nano-regions[13,14]. It appears that this family of compounds is prone for some degree of site-disorder[6], which could facilitate the development of such nano-regions. There is a possibility of an intimate relationship between such electric dipoles and short-range magnetic-order (in other words, MDE effect) in this situation as well. While it is desirable to extend the studies of reference 5 on the Nd compound to temperatures higher than 30 K, this effect could be weak in the Nd compound, considering that Nd carries much less magnetic-moment than Gd. In this connection, it may be recalled that we have provided ample evidence for the role of such short-range magnetic correlations on ME effect in other systems[14,16]. We have also obtained the curves in the presence of several magnetic fields (10, 30, 50, 80 and 140 kOe) and we did not observe any significant change in these high temperature features. To demonstrate this, we have shown the curves for $H$= 80 kOe (figures 6 c-d). Finally, a careful look at the tanδ curves (figures 6b and 6d) at high frequencies reveals the existence of at least one shoulder (around 125 K for 100 kHz) in addition to the peak (around 150 K). This could imply multiple relaxation rates[17].

Figure 7 presents the temperature dependent Raman spectra. As mentioned earlier, the structure of this compound belongs to *Pbam* symmetry, having Gd, Ti in 4g, 4h Wyckoff sites, respectively, while the O atoms occupy the 8i, 4g, 4h, 4e sites. The spectra show a large number of vibrational modes. In fact the mode classification predicts the system to have 48 (=13$A_g$+13$B_{1g}$+11$B_{2g}$+11$B_{3g}$) Raman active modes. In the absence of polarization selective measurements, we have used the isostructural compound $R_2TiO_5$ as a guide to assign vibrational modes to the spectral features[18]. The low frequency features at about 250 cm$^{-1}$ are the external modes that are related to the motion of Gd/Cr atoms and the higher frequency features are related to the internal features of the $TiO_5$ group. The features near 500 cm$^{-1}$ and 800 cm$^{-1}$ are associated with the bending and stretching modes of the $TiO_5$ group respectively. In addition to the sharp vibrational features, below 150 K, one observes gain in spectral weight in the frequency range 150 to 400 cm$^{-1}$. It appears as a broad asymmetric-peak-like feature over which the sharp vibrational features ride. The intensity variation of this peak with temperature is non-monotonic (see the inset of figure 7); it starts to build up below 150 K and then begins to lose its weight below 20K. In the entire temperature range, the vibrational modes show no noteworthy change. We therefore conclude that the asymmetric spectral feature does not owe its origin to phonon vibration, and we speculate that it could be magnetic in its origin.

In summary, we have investigated magnetic and magnetodielectric behaviour of a new member in another family RCrTiO$_5$ that belongs to the same structure as well-known magnetoelectric system, RMn$_2$O$_5$, in which geometrical frustration plays a role on multiferrocity. We have made many interesting observations. The results suggest that there is no clear-cut evidence for long range magnetic ordering in GdCrTiO$_5$ down to 2 K. Though there are some observations near about 11 K that could be attributable to weak Cr magnetic ordering, this requires further confirmation. Nevertheless, the results unequivocally establish that the magnetic ordering from Cr in this compound is suppressed with respect to that in the Nd analogue. It is unusual that $T_N$ value for the Nd analogue[8] is high (21 K), despite the fact that, as well-known, the magnetic moment on Nd is much smaller compared to that on Gd. We believe that the enhanced $T_N$ value for the Nd case arises from the competing role of 4f-radial extension, which is known in the rare-earth literature to increase as one moves away from Gd towards lighter rare-earths, thereby resulting in enhanced hybridization[19]. We therefore tend to argue that Nd 4f (by way of hybridization) has a role to determine the magnetism of Cr, in contrast to original belief proposed long ago[7]. This compound is characterized by magnetodielectric coupling, interestingly without any feature in the absence of magnetic-field. Additional dielectric data at higher temperatures (well above ~ 50 K) suggest the existence of short range electric ordering; this could be associated with an interplay between crystallographic disorder and short-range magnetic order, even at temperatures as high as about 150 K, as inferred from the presence of a broad asymmetric Raman spectral feature as well. There is a possibility of antisite disorder between Cr and Ti crystallographic positions and this high temperature feature may arise because of hopping conductivity between these two sites. Here, we want to note that we have tried to resolve the possibility of anti-site disorder through X-ray diffraction refinement of our sample, which give the best fit without any antisite disorder (within our resolution of this set up). The resistivity of this compound could not be measure around this temperature because of highly insulating nature. Therefore, it is difficult to shade more light on this aspect. Some spectroscopic/synchrotron studies are warranted to explore this feature in more details. It is of interest to focus future studies to explore the role of geometrically frustrated magnetism (as in the case of $RMn_2O_5$) in this family as well.

**Acknowledgement**

The authors thank Kartik K Iyer for his help in many ways during the course of the experiments.

**Figure Captions and Figures:**

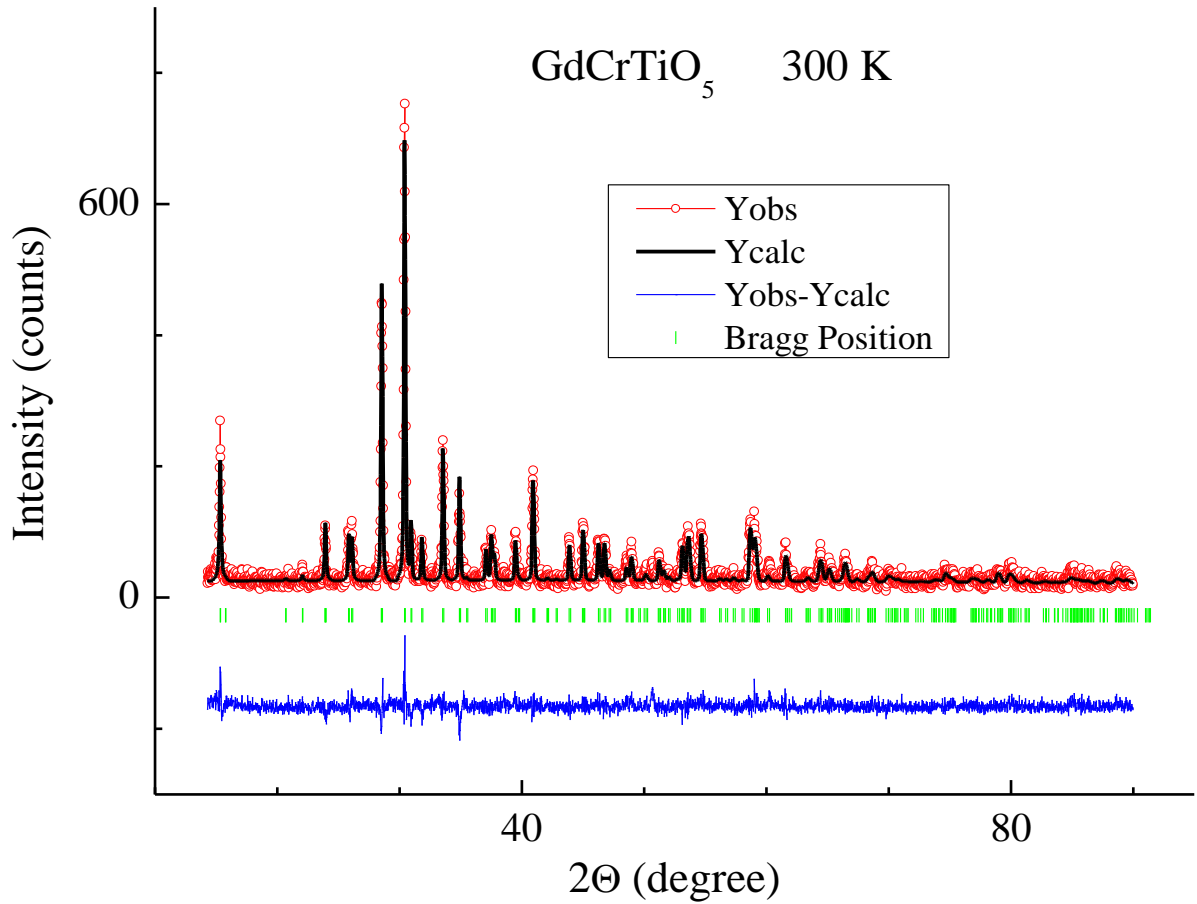

FIGURE 1: Rietveld fit of x-ray diffraction pattern for $GdCrTiO_5$.

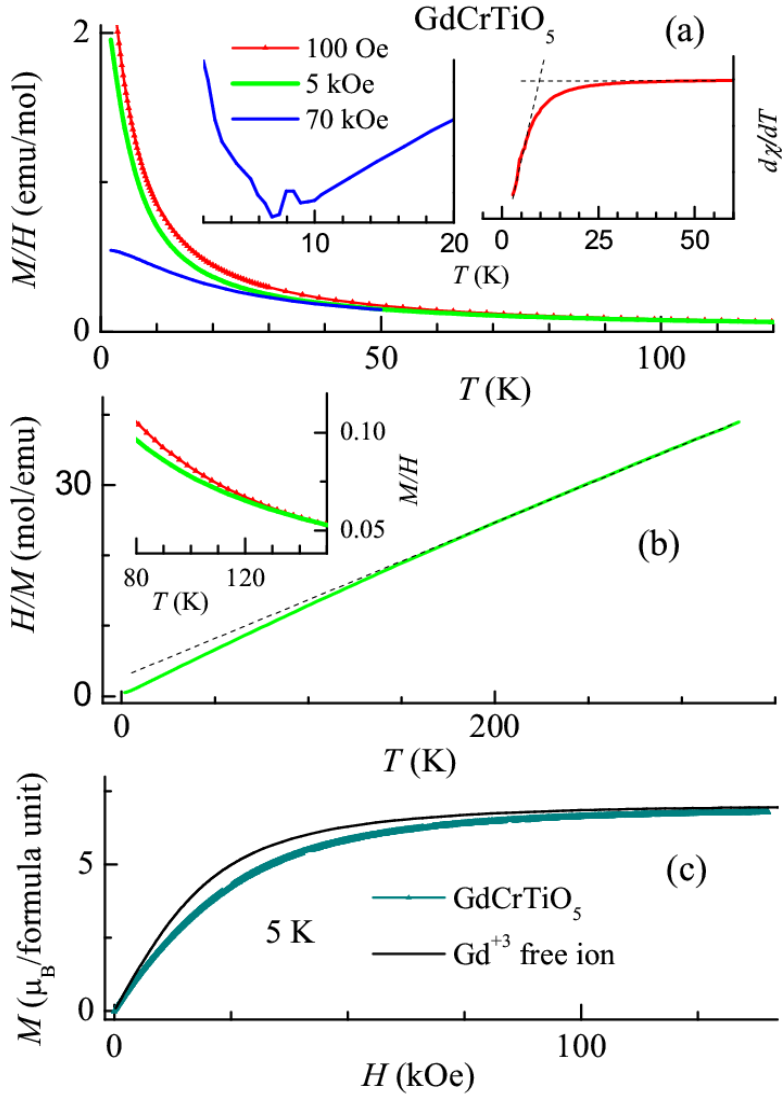

Figure 2: (a) *Dc* magnetic susceptibility ($\chi = M/H$) as a function of temperature (*T*) obtained in the presence of external magnetic fields (*H*= 100 Oe, 5 and 70 kOe) for GdCrTiO$_5$. The curves for zero-field-cooled and field-cooled conditions overlap. (b) Inverse $\chi$ curve obtained in a field of 5 kOe; a line is drawn to show the deviation from the high temperature (>150 K) Curie-Weiss region. (c) Experimentally obtained isothermal magnetization at 5 K for GdCrTiO$_5$ and the theoretically obtained curve for Gd ion from Brillouin function are plotted. Left inset of figure (a): The first derivative of $\chi$ measured in 70 kOe. Right inset of figure (a): The first derivative of $\chi$ measured in 100 Oe and two dashed lines intersecting at $T_N$ are drawn through the data points. Inset of figure (b): zoom out plot of *M/H* vs *T* around 100 K to show a small bifurcation of the curves below 120 K measured in 100 Oe and 5 kOe.

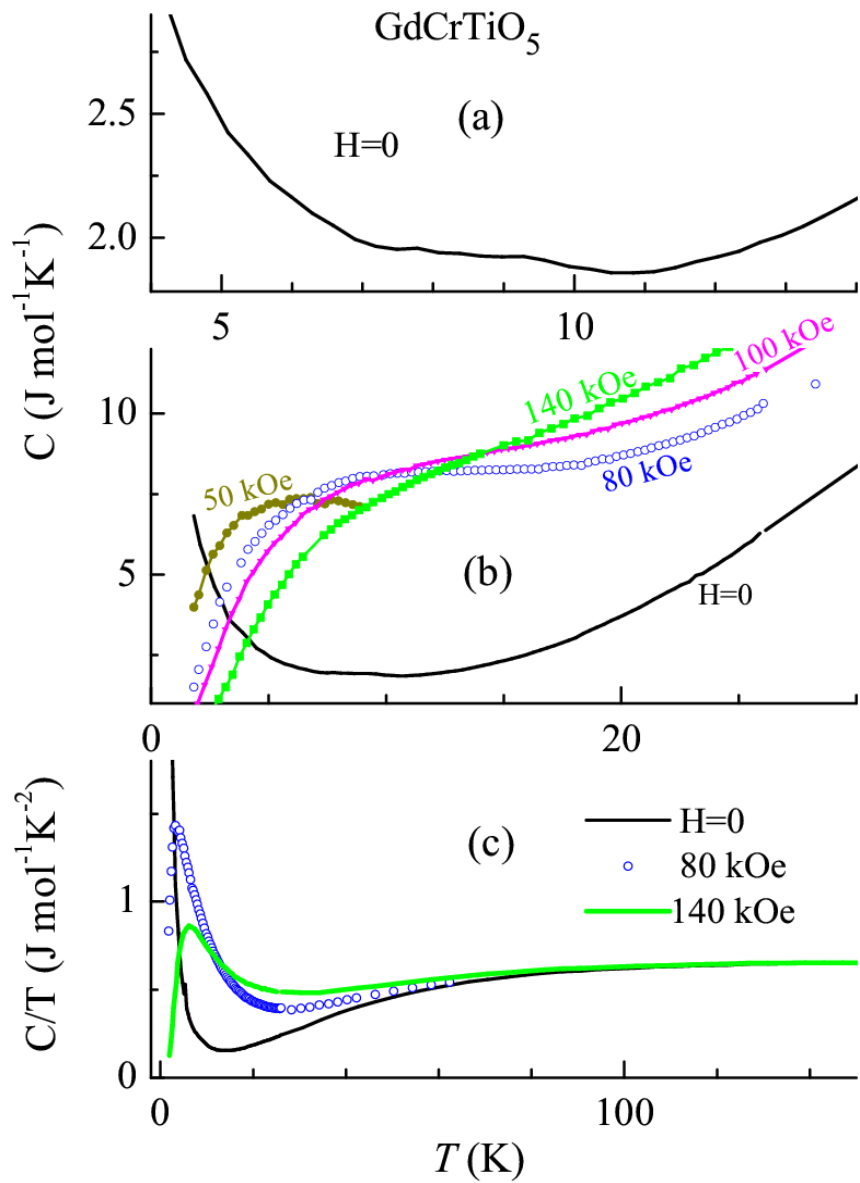

Figure 3: Heat capacity (*C*) as a function of temperature in the range 4-14 K (a) to highlight the feature around 10 K (in the absence of an external field) and (b) in the presence of external fields (0, 50, 80, 100 and 140 kOe). (c) The plot of *C/T* below 150 K at several fields to show that the curves deviate around 10 K.

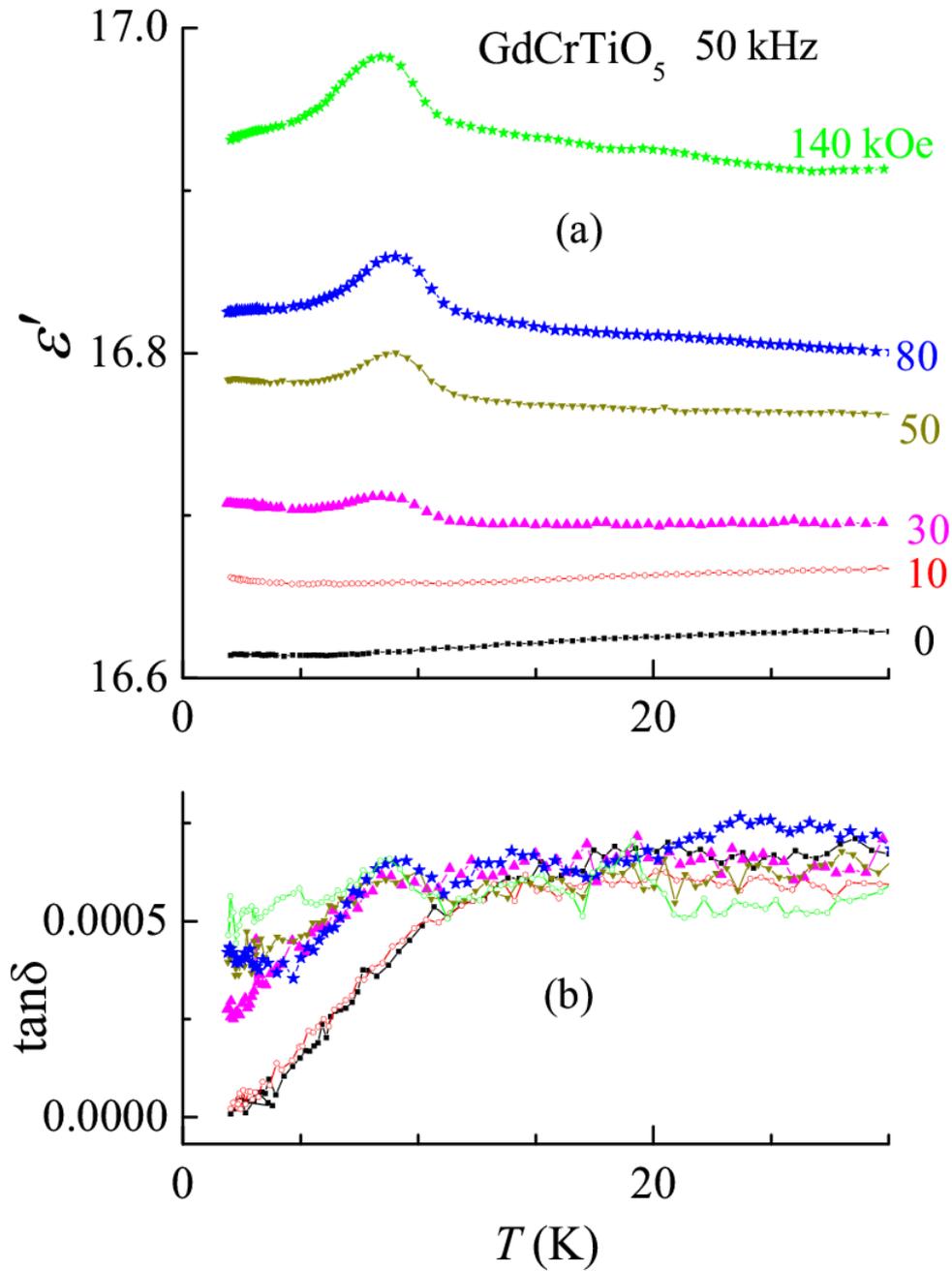

Figure 4: The plot of (a) dielectric constant and (b) loss-factor as a function of temperature (<30K) measured with a frequency of 50 kHz and in fixed magnetic fields for GdCrTiO$_5$ to highlight the development of a peak around 10 K.

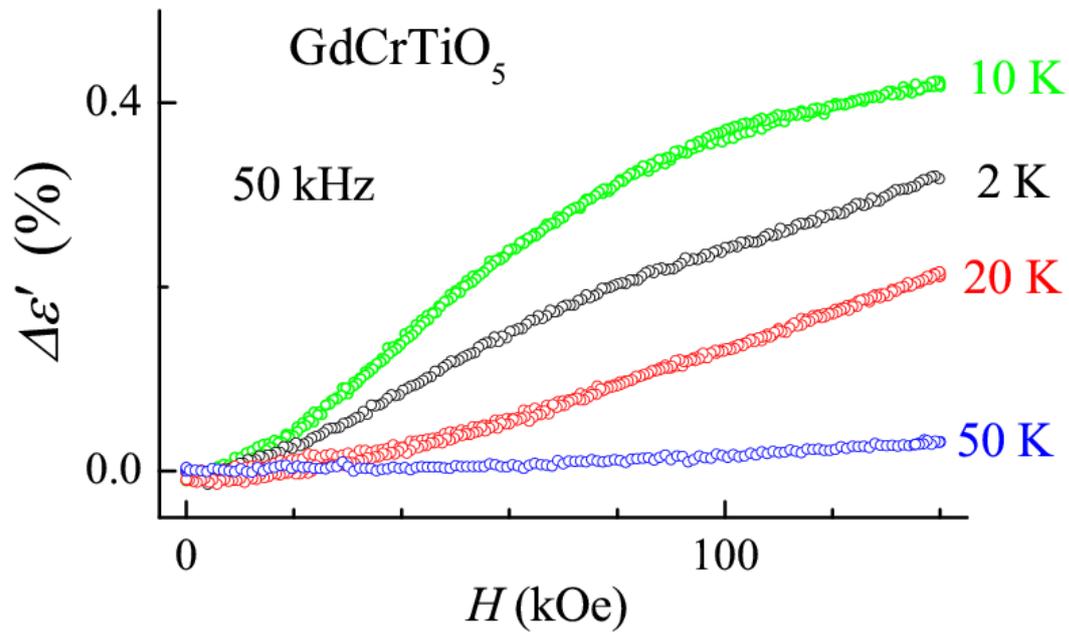

Figure 5: The fractional change in dielectric constant ($\Delta\varepsilon' = (\varepsilon'(H) - \varepsilon'(0))/\varepsilon'(0)$) as a function of magnetic field at 2, 10, 20 and 50 K for GdCrTiO$_5$.

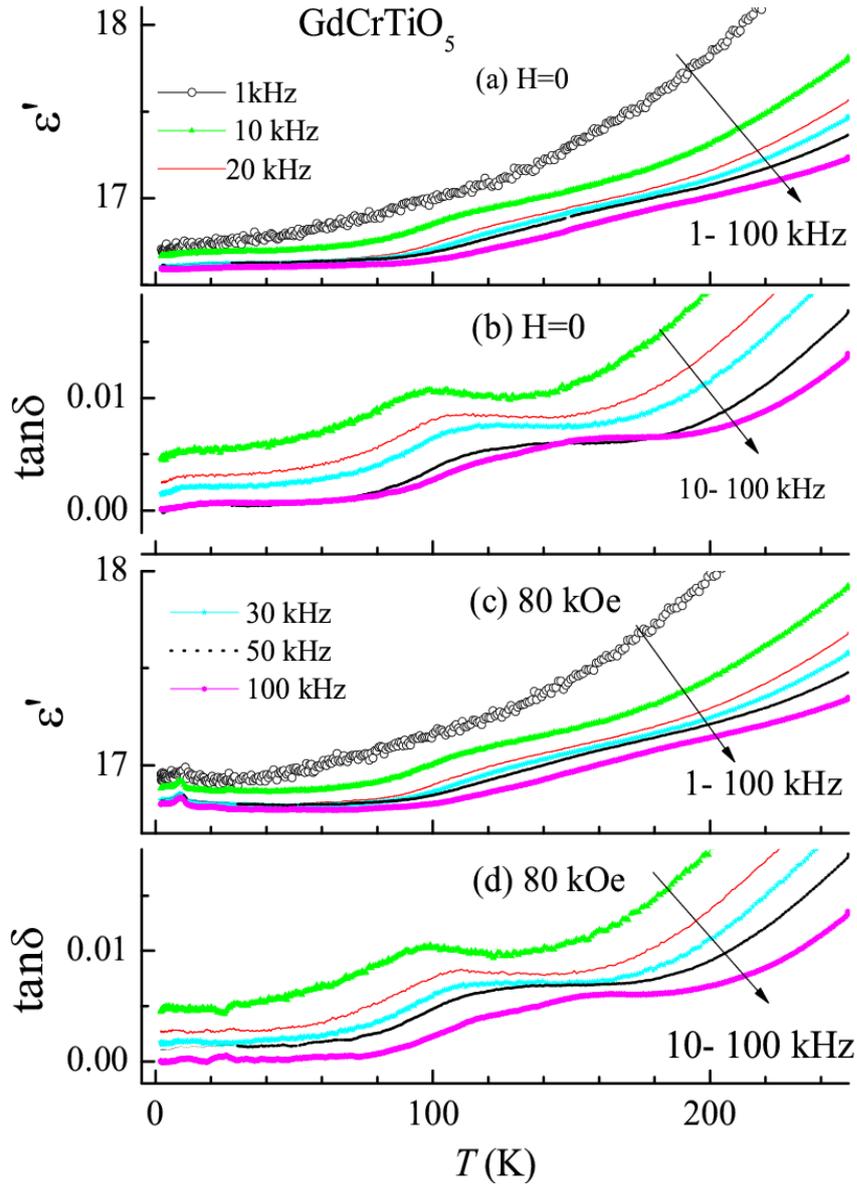

Figure 6: Frequency dependence of dielectric constant and loss-factor below 250 K in (a, b) the absence of magnetic-field, and (**c, d**) 80 kOe for GdCrTiO$_5$.

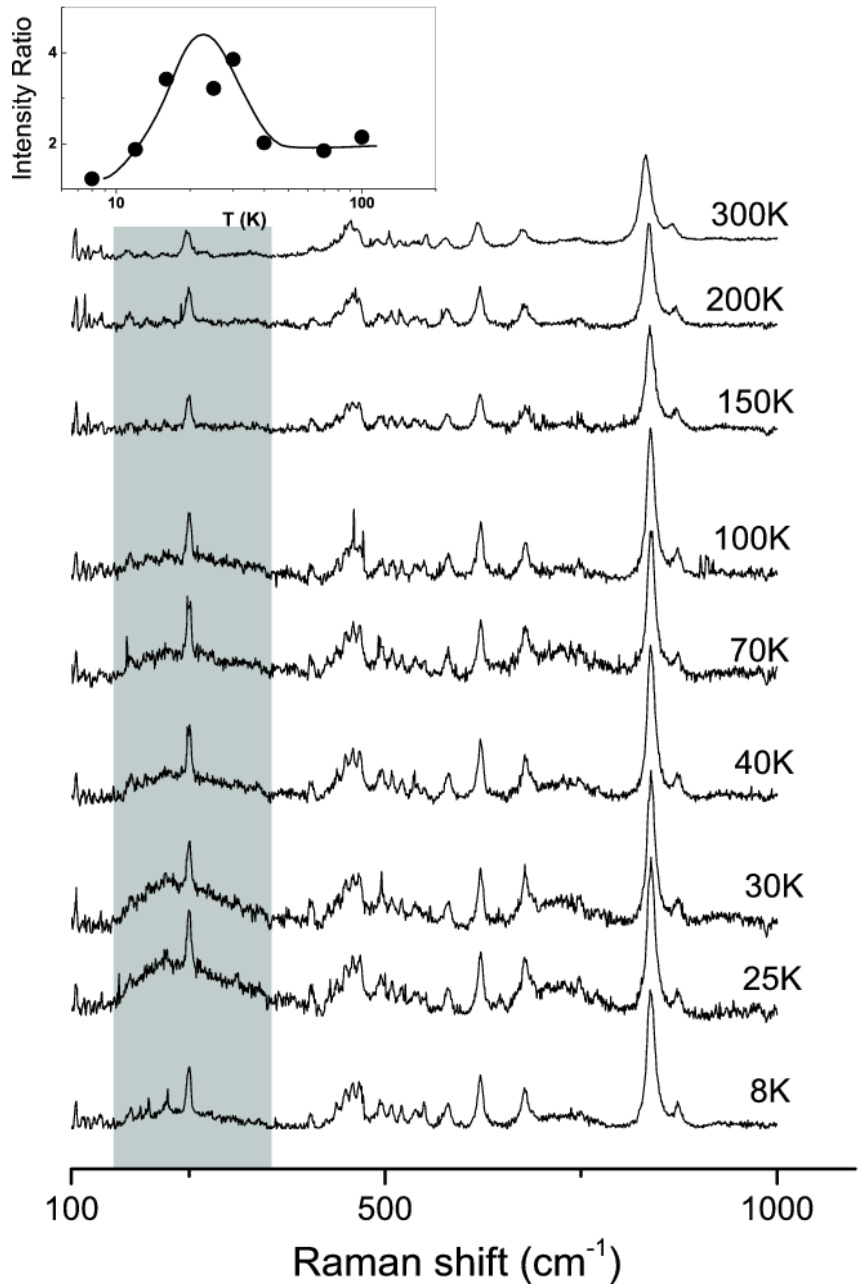

Figure 7: Raman spectra of GdCrTiO$_5$ at select temperatures. The spectral feature in the range 150 to 400 cm$^{-1}$ (marked by shaded band) shows the change in intensity with temperature. Inset shows a non-monotonic variation in the intensity of the broad asymmetric-peak-like feature normalized with the intensity of the stretching mode at 833 cm$^{-1}$ with temperature. The solid line is guide to the eye.